\documentclass[twocolumn]{aastex63}

\usepackage{color}
\usepackage{graphicx}
\usepackage{amsmath, amssymb, bm}
\usepackage{CJK}
\submitjournal{AAS Journals}

\shorttitle{Nightside Clouds of Hot Jupiters}
\shortauthors{Gao \& Powell}

\begin{document}

\title{A Universal Cloud Composition on the Nightsides of Hot Jupiters}

\correspondingauthor{Peter Gao}
\email{pgao8@ucsc.edu}

\author[0000-0002-8518-9601]{Peter Gao}
\altaffiliation{NHFP Sagan Fellow}
\affiliation{Department of Astronomy and Astrophysics, University of California, Santa Cruz, CA 95064, USA}

\author[0000-0002-4250-0957]{Diana Powell}
\affil{Department of Astronomy and Astrophysics, University of California, Santa Cruz, CA 95064, USA}

\begin{abstract}

The day and nightside temperatures of hot Jupiters are diagnostic of heat transport processes in their atmospheres. Recent observations have shown that the nightsides of hot Jupiters are a nearly constant 1100 K for a wide range of equilibrium temperatures (T$_{eq}$), lower than those predicted by 3D global circulation models. Here we investigate the impact of nightside clouds on the observed nightside temperatures of hot Jupiters using an aerosol microphysics model. We find that silicates dominate the cloud composition, forming an optically thick cloud deck on the nightsides of all hot Jupiters with T$_{eq}$ $\leq$ 2100 K. The observed nightside temperature is thus controlled by the optical depth profile of the silicate cloud with respect to the temperature-pressure profile. As nightside temperatures increase with T$_{eq}$, the silicate cloud is pushed upwards, forcing observations to probe cooler altitudes. The cloud vertical extent remains fairly constant due to competing impacts of increasing vertical mixing strength with T$_{eq}$ and higher rates of sedimentation at higher altitudes. These effects, combined with the intrinsically subtle increase of the nightside temperature with T$_{eq}$ due to decreasing radiative timescale at higher instellation levels lead to low, constant nightside photospheric temperatures consistent with observations. Our results suggest a drastic reduction in the day-night temperature contrast when nightside clouds dissipate, with the nightside emission spectra transitioning from featureless to feature-rich. We also predict that cloud absorption features in the nightside emission spectra of hot Jupiters should reach $\geq$100 ppm, potentially observable with the James Webb Space Telescope.

\end{abstract}


\section{Introduction}\label{sec:intro}


Hot Jupiters are likely tidally locked to their host stars due to their close-in orbits, creating a unique laboratory for the investigation of atmospheric heat transport at slow rotation rates, high stellar fluxes, and large longitudinal instellation gradients. An important observable that constrains atmospheric heat transport is the day and nightside temperatures and their difference, as they are controlled by the amount of energy that can be transported from the illuminated dayside to the permanently dark nightside before being radiated to space. Recent \textit{Spitzer} observations of the day and nightside temperatures of hot Jupiters \citep{keating2019,beatty2019} have shown that, while dayside temperatures increase monotonically with planet equilibrium temperature ($T_{eq}$), the nightside temperatures are a largely constant $\sim$1100 K for planets with $T_{eq}$ $<$ 2500 K, leading to increasing day-night temperature differences with increasing $T_{eq}$. Here $T_{eq}$ is defined assuming zero albedo and full heat redistribution. 

Previous studies using three dimensional global circulation models (GCMs) have predicted eastward equatorial jets and sub-to-antistellar flows that carry heat away from the dayside to the nightside, resulting in day-night temperature differences that are dependent on wind and wave speeds and radiative cooling rates \citep[see ][and references therein]{komacek2016,showman2020,parmentier2021}. However, these models often underestimated the day-night temperature difference by overestimating the nightside temperature \citep[e.g.][]{showman2009,wong2016exo,kataria2015,stevenson2017}. One strategy for reconciling model results with the data is to reduce wind speeds and damp out atmospheric waves through a drag process \citep{perezbecker2013,komacek2017,koll2018}, such as turbulence \citep{li2010,fromang2016} and magnetic drag \citep{perna2010,rauscher2013,rogers2014showman,rogers2014komacek}, though the latter process has a strong temperature dependence such that it may not be relevant for cooler planets with $T_{eq}$ $\leq$ 1400 K \citep{rogers2014komacek,koll2018}. Another strategy is to decrease the pressure level probed by observations to reduce the radiative timescale by introducing additional atmospheric opacity, such as by increasing metallicity \citep{showman2009,kataria2015}, though the predicted nightside temperatures still tend to be higher than observed. 

Here we consider the additional opacity of clouds. Observational signatures of clouds on hot Jupiters are ubiquitous, such as muted gas spectral features and scattering slopes in optical and near-infrared transmission spectra \citep{sing2016,barstow2017} and westward-shifted brightness maxima in optical phase curves \citep{demory2013,shporer2015}. GCM studies that allow for cloud formation have shown that the nightsides and western (morning) limbs of hot Jupiters are likely to be cloudy for a large range of $T_{eq}$, assuming thermochemical equilibrium cloud compositions \citep[e.g.][]{parmentier2016}, while the daysides are mostly cloud-free. Indeed, both \citet{keating2019} and \citet{beatty2019} hypothesized that their findings of low, uniform nightside temperatures could be due to the existence of an optically thick cloud deck that permeates the nightsides of all hot Jupiters with $T_{eq}$ $<$ 2500 K, such that the brightness temperature is the atmospheric temperature at the top of the cloud deck, which in turn is tied to the condensation temperature of this ``universal'' cloud species. \citet{parmentier2021} showed using a GCM and a prescribed nightside cloud deck that the brightness temperature observations can be explained by a combination of decreasing radiative timescales with increasing $T_{eq}$ leading to a more constant nightside temperature and clouds pushing up the nightside photosphere to higher altitudes and lower temperatures. However, equilibrium condensation models predict a variety of cloud species forming at the temperatures of hot Jupiter atmospheres, from cooler sulfide clouds to hotter silicates, iron, and aluminum and titanium oxide clouds \citep{visscher2010,morley2012}, suggesting multiple cloud decks with temperature-dependent vertical locations. In addition, the cloud top pressure is not controlled solely by the cloud base pressure, but by the extent to which cloud particles can be lofted to higher altitudes, determined by the complex interaction of microphysical processes like mixing, sedimentation, and particle growth \citep{gao2018b}.

In this letter, we use the one dimensional aerosol microphysics model CARMA (Community Aerosol and Radiation Model for Atmospheres) to compute the cloud distributions and brightness temperatures on the day and nightsides of hot Jupiters for a range of $T_{eq}$. In our previous work \citep{gao2020}, we showed that the differing nucleation energy barriers of the predicted exoplanet condensates led to the dominance of silicates in the total cloud opacity of hot Jupiters for 1000 K $<$ $T_{eq}$ $<$ 2000 K, qualitatively consistent with interpretations of the nightside brightness temperature data. However, that study was conducted using 1D globally averaged temperature-pressure profiles, whereas here we treat the day and nightsides separately. This work extends \citet{powell2018}, which also used CARMA to compute cloud distributions on the day and nightsides, by including cloud species beyond TiO$_2$ and MgSiO$_3$. We also expand upon \citet{powell2019}, which used CARMA to investigate cloud distributions at the limbs of hot Jupiters. Our work complements past studies that use GCMs with simpler cloud parameterizations to investigate the global distribution of clouds on hot Jupiters and cloud feedback \citep[e.g.][]{parmentier2016,roman2019,lines2019,roman2021,parmentier2021} by focusing instead on cloud microphysics and how it shapes the cloud composition and vertical extent. We also differ from GCM studies of hot Jupiters that consider detailed cloud microphysics \citep[e.g.][]{lee2015,helling2016,lines2018}, as we are considering a wide range in $T_{eq}$ in order to reproduce the observed trend in day-night temperature differences instead of modeling individual exoplanets.



In ${\S}$\ref{sec:model}, we give an overview of our model background atmospheres, CARMA, and our calculation of the day and nightside brightness temperatures. We compare our computed brightness temperatures to data in ${\S}$\ref{sec:results} and describe the physical mechanisms at work. Finally, in ${\S}$\ref{sec:discussion} we discuss observational tests of our model and how our results contrast with those of previous works. 


\section{Model}\label{sec:model}

\subsection{Atmospheric Thermal Structure}\label{sec:thermal}

\begin{figure*}[hbt!]
\centering
\includegraphics[width=0.9 \textwidth]{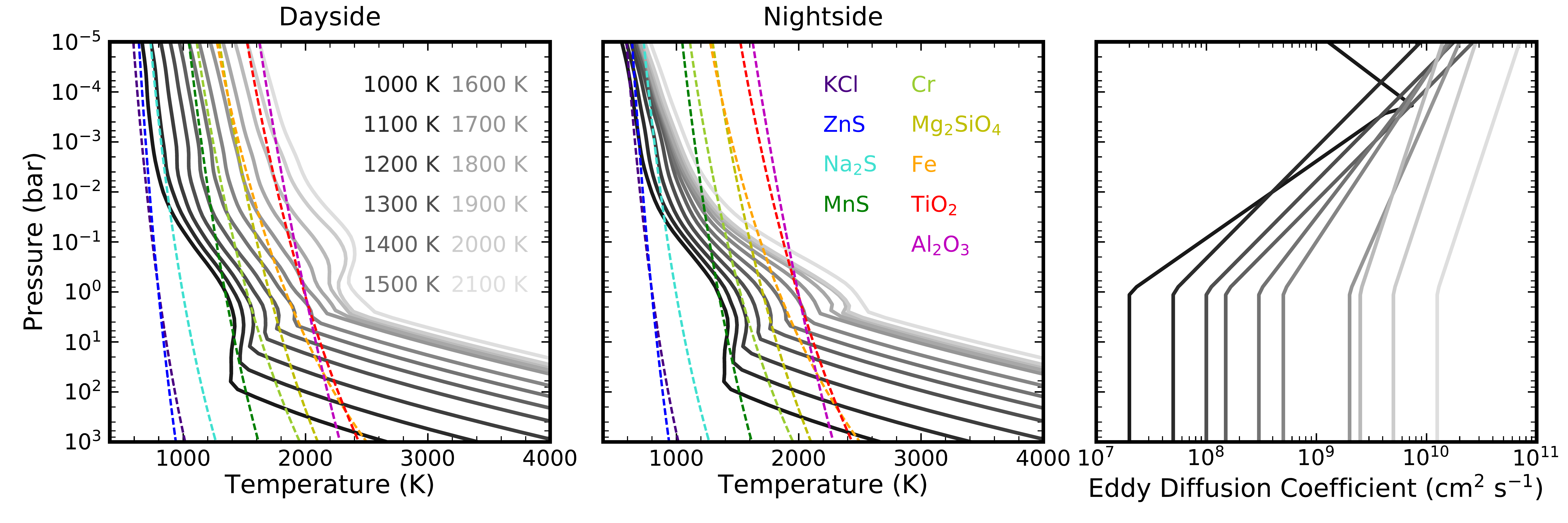}
\caption{Temperature-pressure profiles for the daysides (left) and nightsides (middle) of our model hot Jupiters compared to condensation curves of the considered cloud species. The right panel shows the eddy diffusion coefficient profiles used for both the day and nightsides.  }
\label{fig:tpkzz}
\end{figure*}

We generate separate temperature-pressure (TP) profiles for the day and nightsides of hot Jupiters from the grid of cloudless GCM simulations of \citet{parmentier2016}. The model planets all have a gravity of 10 m s$^{-2}$ at the 1 bar pressure level and solar metallicity atmospheres without TiO/VO. The radii of the planets at 1 bar are computed using a planetary interior model \citep{thorngren2016} and are used in calculating the profile of gravitational acceleration in the atmosphere given the fixed 10 m s$^{-2}$ gravity at 1 bar. We consider planets with $T_{eq}$ from 1000 to 2100 K in 100 K steps. Higher $T_{eq}$ models are not considered due to the onset of more complex atmospheric processes such as H$_2$ dissociation and magnetically coupled circulation \citep{tan2019ultra,rogers2014komacek}. TP profiles are sampled at each longitude and latitude point for a given 3D model atmosphere, and the profiles within 20$^{\circ}$ of the sub- and antistellar points are averaged to produce the day and nightside TP profiles for each instance (Figure \ref{fig:tpkzz}), respectively.  

We modify the averaged TP profiles by setting the radiative convective boundary (RCB) to those computed in \citet{thorngren2019} to take into account the observed degree of inflation of hot Jupiters. We assume an adiabatic gradient below the new RCB described by \citep{parmentier2015},

\begin{equation}
\label{eq:adiabat}
\left ( \frac{d\ln{T}}{d\ln{P}} \right )_S = 0.32 - 0.1 \left ( \frac{T}{3000\, {\rm K}} \right )
\end{equation}

\noindent where $T$ and $P$ are temperature and pressure, respectively. Our modification implies a higher internal heat flux than previously assumed and drastically reduces the pressure level of the RCB, from hundreds of bars to 1-10 bars for the hotter ($T_{eq}$ $>$ 1200 K) models. Importantly, this eliminates the deep radiative region and therefore the deep cold trap hypothesized by \citet{parmentier2016} to sequester TiO and silicate clouds below the photosphere, allowing these condensates to impact emission observations to much lower $T_{eq}$ \citep{powell2018}. In addition, the shallower RCB and greater internal heat flux could affect the circulation pattern and thus change the day- and nightside TP profiles. However, while several recent works have considered higher internal heat fluxes \citep{carone2020,beltz2021,steinrueck2021}, none have isolated the impact of changing internal heat flux on the circulation pattern, and therefore we do not consider its influence on the TP profile here beyond the shallower RCB. 

A single eddy diffusion coefficient ($K_{zz}$) profile is assumed for each set of day and nightside TP profiles for a given $T_{eq}$ (Figure \ref{fig:tpkzz}) to simulate atmospheric mixing. Each profile is constructed by fitting power law functions of the atmospheric pressure level to the $K_{zz}$ profiles computed by the GCM through tracer transport \citep{parmentier2013,powell2019}. The increase in $K_{zz}$ in our model with increasing $T_{eq}$ is similar to behavior seen in other recent studies that seek to quantify $K_{zz}$ using GCMs \citep{komacek2019}.  We assume a constant $K_{zz}$ below 1 bar equal to the 1 bar $K_{zz}$ value for all cases to avoid non-convergence issues due to low $K_{zz}$ and to take into account the shallower radiative convective boundary. This may result in an underestimation of the $K_{zz}$ in the convective zone if it were well-represented by mixing length theory, which predicts values closer to 10$^9$--10$^{11}$ cm$^2$ s$^{-1}$ for the $T_{eq}$'s considered here \citep{ackerman2001,gao2020}. We evaluate the impact of this possible issue in ${\S}$\ref{sec:results}.


\subsection{Cloud Microphysics}\label{sec:micro}

We simulate cloud distributions for our model atmospheres using CARMA, which computes vertical and size distributions of aerosol particles by solving the discretized aerosol continuity equation in a bin scheme, taking into account rates of aerosol nucleation, condensation, evaporation, coagulation, and transport \citep{turco1979,toon1988,jacobson1994,ackerman1995}. We refer the reader to the appendix of \citet{gao2018a} for a complete description of CARMA. 

We use the same model setup as in \citet{gao2020}. Briefly, we consider the homogeneous nucleation of TiO$_2$, Fe, Cr, and KCl and the heterogeneous nucleation of Al$_2$O$_3$, Fe, Mg$_2$SiO$_4$, Cr, MnS, and Na$_2$S on the pure TiO$_2$ particles, and ZnS on the pure KCl particles. Heterogeneous nucleation leads to layered particles in CARMA, where the core is the original cloud condensation nucleus (CCN) and the mantle is the nucleating species. The desorption energy of a condensate molecule on the surface of the CCN is set to 0.5 eV as an average between higher desorption energies due to chemical bond formation and lower desorption energies due to weaker van der Waals interactions. The contact angle $\theta_c$ between the condensate germ and the CCN is determined by 

\begin{equation}
\label{eq:contangle}
\cos{\theta_c} = \frac{\sigma_C - \sigma_{xC}}{\sigma_x}
\end{equation}

\noindent where $\sigma_C$ is the CCN material surface energy, $\sigma_x$ is the condensate surface energy, and $\sigma_{xC}$ is the interfacial energy between the condensate germ and the CCN, assumed to be zero due to lack of measured data; non-zero (positive) $\sigma_{xC}$ leads to higher $\theta_c$ and lower nucleation rates. 

The model atmosphere is initialized with no cloud particles and condensate vapor only present at the highest pressure level. The mixing ratios of the condensate vapors are their solar metallicity values for all but TiO$_2$ and KCl, which must compete for their limiting element (Ti and K, respectively) with other species (TiO and K and KOH, respectively), yielding lower mixing ratios computed using the thermochemical equilibrium model GGChem \citep{woitke2018}. Upon initialization, the condensate vapors are mixed upwards in the atmosphere and may reach supersaturation, at which point they may nucleate. The top boundary condition is set to be zero-flux, while the bottom boundary condition is set to be the initial mixing ratios for the condensate vapors and zero mixing ratio for the cloud particles, as the bottom boundary pressure is selected to prevent supersaturation (and therefore nucleation) of any of the condensate vapors. 

The cloud optical depths, single scattering albedos, and asymmetry parameters for the different cloud species are calculated using \texttt{pymiecoated} \citep{leinonen2016}, taking into account the contributions from the core and mantle. We refer the reader to the Supplementary Information of \citet{gao2020} for the sources of the refractive indices for the condensates used here, as well as the material properties mentioned in the previous paragraphs. 


\subsection{Brightness Temperatures}\label{sec:btemp}

We calculate the brightness temperature spectrum using a 1D thermal structure code that computes TP profiles in radiative-convective-thermochemical equilibrium \citep{mckay1989,marley1996,fortney2005,saumon2008}. However, we do not consider a radiative-convective equilibrium TP profile; instead, we use the code to compute the atmospheric composition assuming thermochemical equilibrium along the day and nightside TP profiles generated in ${\S}$\ref{sec:thermal}, taking into account condensation. The TP profile, gas composition, and cloud optical properties are then used by the code in a radiative transfer calculation to arrive at the brightness temperature as a function of wavelength. Cloud feedback on the TP profile is not taken into account, and as such the thermal structure and cloud distribution are not self-consistent. This is unlikely to affect our results for the nightside models, but will likely impact our dayside models more strongly, as we will discuss in ${\S}$\ref{sec:discussion}.

\section{Results}\label{sec:results}

\begin{figure*}[hbt!]
\centering
\includegraphics[width=0.9 \textwidth]{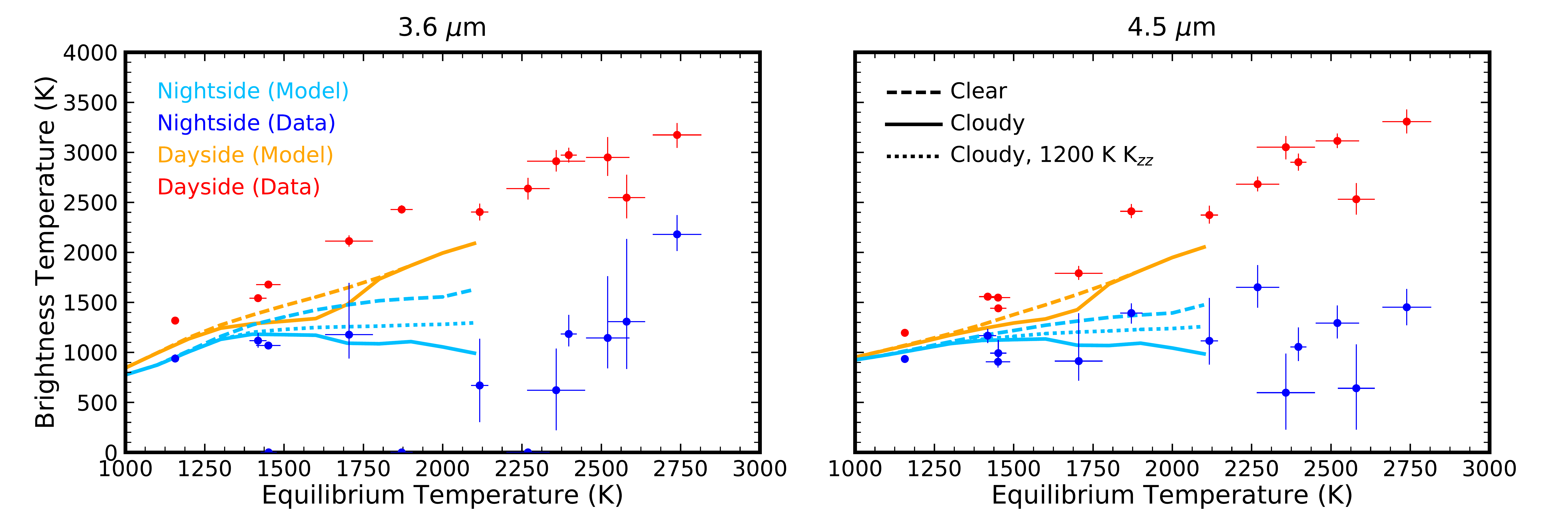}
\caption{Comparison of modeled day- (orange) and nightside (light blue) brightness temperatures with (solid) and without (dashed) clouds at 3.6 (left) and 4.5 $\mu$m (right) with observations (dayside: red; nightside: blue) from \citet{beatty2019}. The dotted light blue curve shows the cloudy nightside brightness temperature when the $K_{zz}$ for all cases with $T_{eq}$ $>$ 1200 K is set to that of the $T_{eq}$ = 1200 K case. We omit the comparison of our results to the day- and nightside temperatures of \citet{keating2020}, as they present the effective temperatures computed from the brightness temperatures.} 
\label{fig:beattycompare}
\end{figure*}

The formation of optically thick clouds near the photospheres of the nightsides of our model hot Jupiters lowers their brightness temperatures in the \textit{Spitzer} 3.6 and 4.5 $\mu$m bands to $\leq$1100 K for $T_{eq}$ $>$ 1200 K, inline with the observations (Figure \ref{fig:beattycompare}). Clouds are situated too deep in the atmosphere to affect the brightness temperature for lower $T_{eq}$'s. In comparison, the brightness temperatures of the cloudless nightside models in the same bands increase with $T_{eq}$, resulting in higher brightness temperatures compared to the data.


\begin{figure}[hbt!]
\centering
\includegraphics[width=0.5 \textwidth]{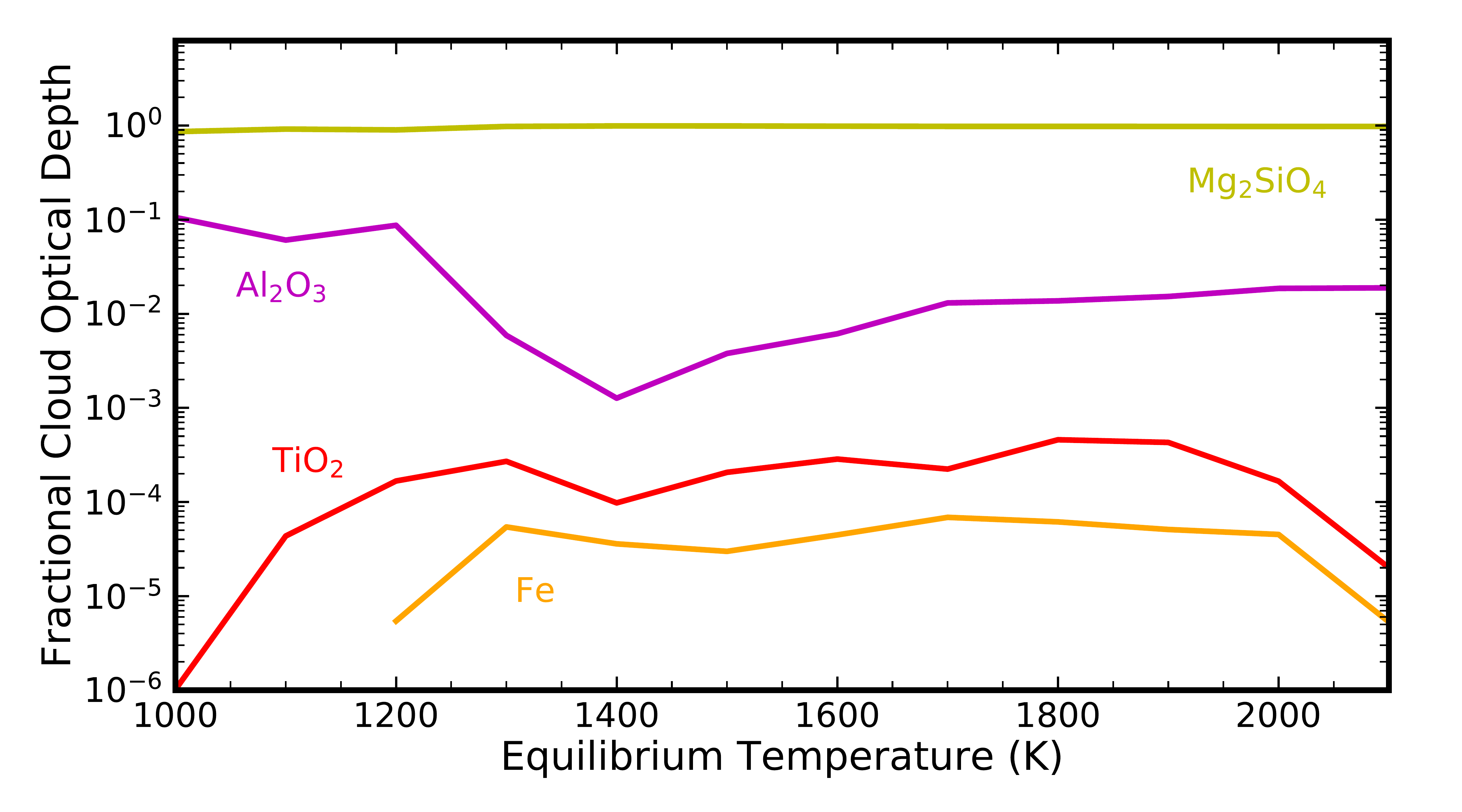}
\caption{Fractional contributions to the cloud optical depth from our considered condensates at the pressure levels probed by the 4.5 $\mu$m \textit{Spitzer} band. The results for the 3.6 $\mu$m band is nearly identical. All condensates aside from those shown have negligible optical depth.}
\label{fig:cloudcomp}
\end{figure}

\begin{figure*}[hbt!]
\centering
\includegraphics[width=0.9 \textwidth]{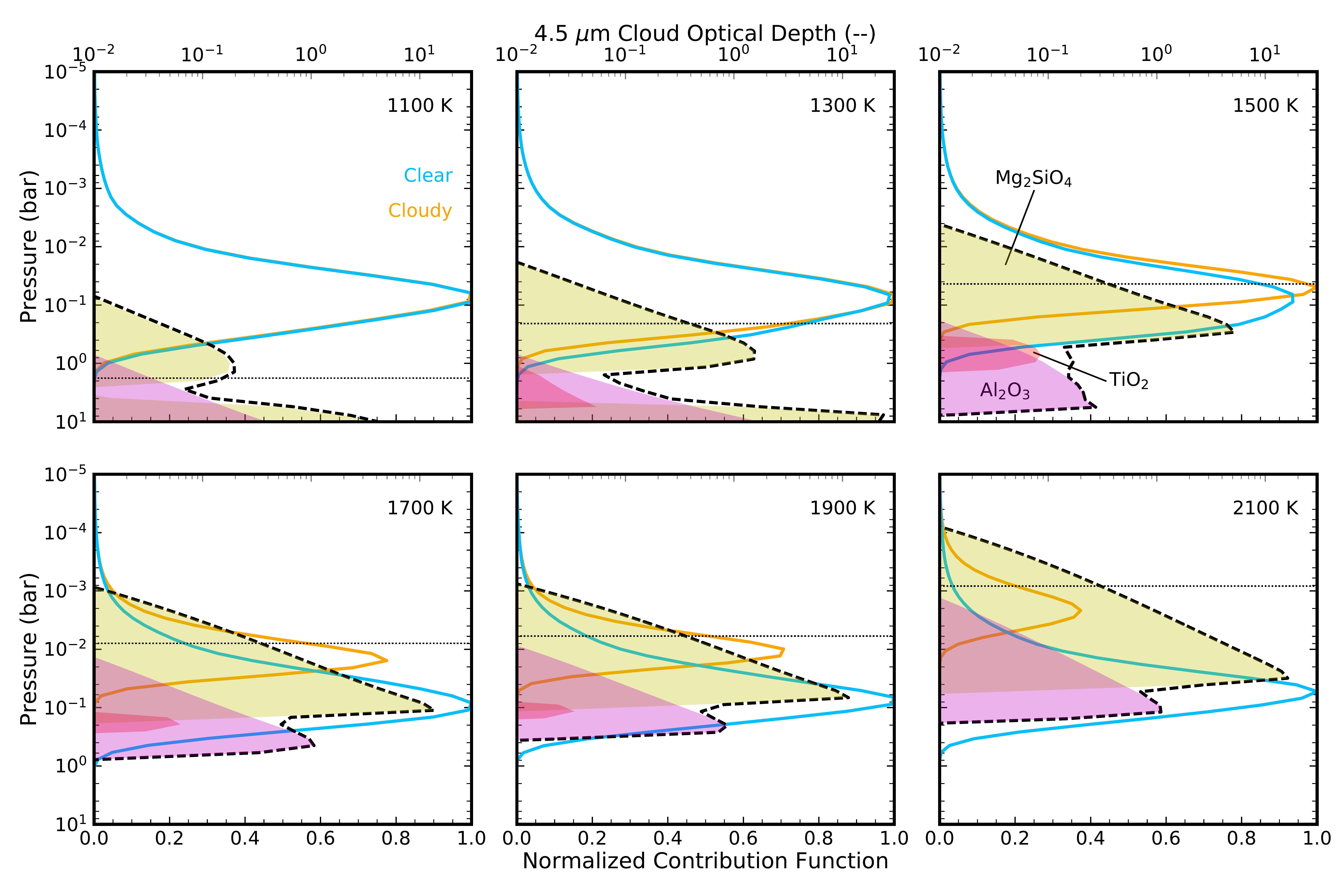}
\caption{Contribution functions of the 4.5 $\mu$m \textit{Spitzer} band for the nightsides of our model hot Jupiters with (orange) and without (blue) clouds, normalized to the cloudless curve. The results for the 3.6 $\mu$m band is nearly identical. The cloud optical depth at each layer is shown in the shaded areas for Mg$_2$SiO$_4$ (yellow), TiO$_2$ (red), and Al$_2$O$_3$ (magenta), with the total optical depth at each layer plotted in the dashed curve. The pressure level where the cumulative cloud optical depth (integrated from the top of the atmosphere downwards) equals 1 is marked by the horizontal dotted line. The contribution function does not peak at the dotted line because the radiative transfer calculation takes into account the thermal scattering due to the large single scattering albedo of the silicate clouds. }
\label{fig:cfnight}
\end{figure*}

Our results show the natural consequence of the slow rise in nightside temperatures with increasing $T_{eq}$ due to decreasing radiative timescales combined with the dominance of silicate clouds in the nightside atmospheric opacity for 1000 K $<$ $T_{eq}$ $<$ 2000 K (Figure \ref{fig:cloudcomp}). As in our previous work \citep{gao2020}, the dominance of silicate clouds is due to the higher abundance of Mg and Si in a solar composition gas and the low surface energies of silicate melts compared to the surface energies of Fe and sulfides, also abundant in vapor form. The low surface energies lead to small contact angles and thus high nucleation rates, resulting in silicate dominance. Thus, the cloud base on the nightsides of hot Jupiters is inexorably tied to the silicate condensation temperature, as hypothesized by \citet{beatty2019,keating2019}, with increasing $T_{eq}$ pushing the silicate cloud to higher altitudes and cooler temperatures (Figure \ref{fig:cfnight}). The modeled brightness temperature is then controlled by the location within the silicate cloud where the cumulative cloud optical depth reaches unity. This is determined by the cloud vertical extent, which in turn depends on the vertical mixing rate and therefore the $K_{zz}$ profile. In our nominal set of models with $K_{zz}$ increasing with $T_{eq}$, the increasing rate of mixing compensates for the decreasing sedimentation timescale as the silicate cloud is pushed to higher altitudes. This leads to vertically extended clouds at the highest $T_{eq}$'s that actually results in a slight downward trend in the predicted brightness temperature with increasing $T_{eq}$ (Figure \ref{fig:beattycompare}). 

We test the sensitivity of our results to how $K_{zz}$ varies with $T_{eq}$ by running an alternative set of models where the $K_{zz}$ for all cases with $T_{eq}$ $>$ 1200 K are set to those at 1200 K. The difference in $K_{zz}$ at 1 bar between the nominal models and these models are $\sim$2 orders of magnitude for the hottest cases (Figure \ref{fig:tpkzz}). These alternative models produce a brightness temperature vs. $T_{eq}$ trend inbetween that of the nominal models and the cloudless nightside models (Figure \ref{fig:beattycompare}), as expected from more vertically compact clouds resulting from lower mixing rates. Interestingly, this alternative trend is flatter than our nominal trend, and appears to fit the data about as well as the nominal trend. As such, we conclude that varying $K_{zz}$ by 1--2 orders of magnitude does not strongly impact our results. This is in line with previous work that showed that the steady-state cloud properties in this model are insensitive to increasing or decreasing $K_{zz}$ by a factor of 5 below 1 bar \citep{powell2019}.

We expect that the dominance of silicate clouds  minimizes the impact of the possible underestimation of the $K_{zz}$ in the convective zone (${\S}$\ref{sec:thermal}). In our model, eddy diffusion is vital for controlling the cloud mass through the rate of replenishment of condensate vapor to the cloud forming region in addition to the vertical extent of the cloud. As such, the values of $K_{zz}$ at depth is not nearly as important as those at and above the cloud base. As shown in Figure \ref{fig:tpkzz}, the silicate cloud base on the nightside lies at pressures $<$ 1 bar for all $T_{eq}$ $\geq$ 1400 K, which is the same range of $T_{eq}$ where the impact of nightside clouds becomes pronounced (Figure \ref{fig:cfnight}). Therefore, our results should not be significantly affected by the $K_{zz}$ profile at pressures greater than 1 bar. However, second order effects may exist, as silicate cloud particles use TiO$_2$ cloud particles as CCN, and the TiO$_2$ cloud base lies at higher pressures. A more rigorous evaluation of the validity of our assumptions regarding the impact of the deep $K_{zz}$ value would require 3D atmospheric circulation models of hot Jupiters that extend deeper than the RCB \citep[e.g.][]{carone2020}, which is beyond the scope of this letter.

In addition to silicates, Al$_2$O$_3$, TiO$_2$, and Fe clouds also form on the nightside, though they are much less optically thick at the near-infrared photosphere (Figures \ref{fig:cloudcomp} and \ref{fig:cfnight}). In particular, Fe clouds tend to form deeper in the atmosphere and are restricted to a thin layer near the cloud base. This is because the large surface energy of iron leads to low nucleation rates (both homogeneous and heterogeneous), resulting in large mean particle radii and therefore a vertically compact cloud. Fe vapor is depleted by cloud formation with respect to solar by a factor of 10 for the $T_{eq}$ = 2100 K case, with greater depletion factors for lower $T_{eq}$ cases due to the cooler temperatures and thus higher supersaturations driving higher nucleation rates. Nightside Fe condensation has been proposed to explain the asymmetric atomic Fe absorption of the ultra-hot Jupiter WASP-76b in transmission \citep{ehrenreich2020}, though the actual factor of depletion has not yet been determined. While WASP-76b is much hotter than the planets considered here, its nightside could be cool enough for Fe and other clouds to form due to a slow down of the circulation from magnetic drag of the dayside atmosphere, which is hot enough to ionize atomic species \citep{komacek2016}. However, without considering wind speeds and the zonal transport of cloud and vapor we cannot make a robust prediction for the actual nightside Fe vapor depletion. 

In contrast to the nightside models, the cloudy dayside models significantly underestimate the dayside brightness temperatures (Figure \ref{fig:beattycompare}), and predict a change in the slope of the brightness temperature versus $T_{eq}$ relation that is not observed. This change in slope is due to the same effect that generates the constant nightside temperature: an optically thick silicate cloud forming at the photosphere. However, because the dayside temperature rises much more quickly than the nightside (Figure \ref{fig:tpkzz}), the dayside clouds in our model rapidly dissipate for $T_{eq}$ $\geq$ 1700 K. This results in a return to the cloudless model track, which better replicates the observations, though they still underestimate them by 100-200 K.

\section{Discussion}\label{sec:discussion}


\subsection{Impact of Cloud Feedback}

We do not consider the impact of cloud formation on the atmospheric thermal structure. Previous GCM studies that investigated cloud feedback showed that clouds could have multiple competing effects that are strongly dependent on the vertical, compositional, and size distribution of cloud particles. For example, highly reflective clouds (e.g. silicates and sulfides) on the dayside can cool the atmosphere and reduce global emission, while absorbing clouds (Fe and Al$_2$O$_3$) can heat the atmosphere, leading to thermal inversions, cloud dissipation, and higher dayside fluxes. On the nightside, clouds can limit the nightside emission and cool the nightside atmosphere above the clouds while heating the atmosphere below the clouds. Importantly, nightside clouds also lead to hotter daysides, which must now radiate nearly all of the captured stellar flux. Globally, vertically extended clouds tend to impact the thermal structure more than vertically compact clouds, while the cloud particle size determines the magnitude of cloud radiative forcing \citep{roman2019,roman2021,parmentier2021}. 

In the context of our work, the lack of significant Fe clouds above the photosphere means that, on average, cloud heating on the dayside should be less effective and dayside clouds should be more reflective, though Al$_2$O$_3$ will still impart some absorbance. More importantly, the increased dayside temperatures due to lower nightside fluxes could lead to more optically thin dayside clouds or possibly no dayside clouds at all apart from possible clouds on the western limb, improving the fit to the dayside brightness temperature observations \citep[Figure \ref{fig:beattycompare}; see also][]{roman2021,parmentier2021}. The optical depth profiles of our simulated clouds at near-infrared wavelengths show that the clouds are vertically extended over multiple scale heights above the cloud base: the optical depth drops by roughly a factor of 1/e over a pressure scale height (Figure \ref{fig:cfnight}), consistent with the definition of a vertically extended cloud as expressed in Eq. 2 of \citet{roman2021}. This is due to the dominance of mixing over sedimentation as the primary transport mechanism in our model near the cloud base. However, this does not extend to the top of the modelled domain, shown in the change in slope of the optical depth profile in the upper reaches of the clouds, as above a certain pressure level sedimentation becomes the dominant transport mechanism \citep{parmentier2013}. The particle size distributions that we have computed are similar to those of \citet{powell2019}, showing a broad plateau that extends from submicron to $\geq$10 $\mu$m particles. These results imply that our computed cloud distributions should significantly affect the atmospheric thermal structure. However, it should be noted that changes to the thermal structure should not affect our proposed mechanism for generating the observed low, near-constant nightside temperature -- silicates dominating the nightside cloud composition -- as it is robust to changes in temperature of $\sim$1000 K \citep{gao2020}.


\subsection{Observational Tests and Future Work}

\begin{figure*}[hbt!]
\centering
\includegraphics[width=0.9 \textwidth]{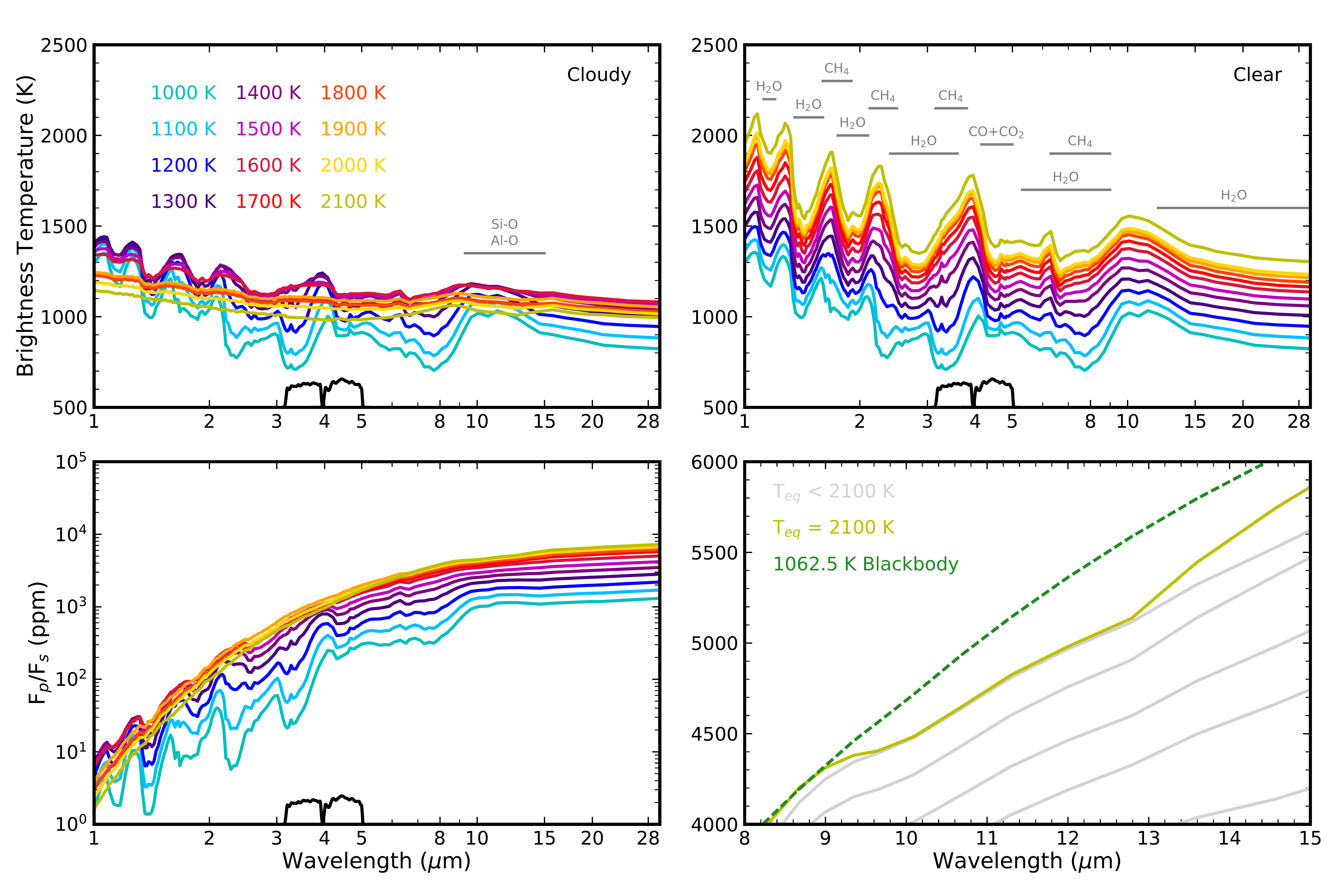}
\caption{(Top) Brightness temperature as a function of wavelength for the nightsides of our simulated hot Jupiters, assuming cloudy (left) and clear (right) atmospheres. Spectral features associated with several major molecules and clouds are indicated. (Bottom left) The planet-to-star flux ratio for the nightsides of our simulated hot Jupiters. We assume the Sun for the host star and use the ASTM E-490 standard reference solar spectrum. (Bottom right) Same as bottom left, but with a focus on thermal IR wavelengths. The case with the largest amplitude cloud spectral features ($T_{eq}$ = 2100 K) is emphasized (yellow) and compared to the planet-to-star flux ratio of a blackbody with a temperature of 1062.5 K and the same radius (green). The \textit{Spitzer} 3.6 and 4.5 $\mu$m filter profiles are shown at the bottom of the top and bottom left panels.}
\label{fig:spec}
\end{figure*}

As nightside clouds form at ever greater altitudes in the atmosphere with increasing $T_{eq}$, we predict that molecular features in the nightside emission spectra of hot Jupiters, including those of H$_2$O, CO, CH$_4$, and CO$_2$ should gradually fade until the spectra become blackbodies with effective temperature $\sim$1100 K. This should occur for $T_{eq}$ $\geq$ 1800 K (Figure \ref{fig:spec}) and at wavelengths as long as 30 $\mu$m due to the large cloud particle radii. However, clouds should eventually become optically thin with increasing $T_{eq}$, as the absolute abundance of condensates decrease with decreasing cloud base pressure. This process could be accelerated by atmospheric heating mechanisms that emerge at T$_{eq}$ $>$ 2100 K, such as the formation of hot stratospheres due to absorption by TiO/VO and atomic metals \citep{fortney2008tiovo,evans2017,lothringer2018heat} and nightside H$_2$ recombination \citep{tan2019ultra}. Therefore, we expect a regime change at some critical temperature (or range of temperatures) in the global thermal structure--and thus day-night temperature difference--of hot Jupiters, likely at 2000 K $<$ $T_{eq}$ $<$ 3000 K. Below this critical temperature, nightsides are cloudy and possess an emission temperature capped at 1100 K, with dayside emission being responsible for the majority of the planet's cooling. Above the critical temperature, nightside clouds dissipate due to some combination of low condensate abundance, stratosphere formation, and H$_2$ recombination heating, significantly increasing nightside emission and reducing the day-night temperature difference \citep{roman2021}. A key prediction of this scenario is that the nightside emission spectrum should transition from a low effective temperature, largely featureless blackbody to one rich with gas spectral features at a much higher flux as planet $T_{eq}$ increases past the critical temperature. A survey of nightside emission spectra at the relevant $T_{eq}$ ranges with e.g. the James Webb Space Telescope (JWST) or Ariel would be needed to test this hypothesis. 

As CARMA is able to resolve the full size distribution of cloud particles, we are able to predict the amplitude of the silicate (Si-O) absorption feature at $\sim$10 $\mu$m, which is strongly dependent on the width of the size distribution and mean particle size \citep{wakeford2015}. Measuring this feature is vital for constraining the dominant cloud composition and particle properties. We predict that the amplitude of the feature should be $\sim$100-200 ppm in terms of planet-to-star flux ratio for the best targets orbiting Sun-like stars (Figure \ref{fig:spec}). This level of precision could be achievable with the MIRI instrument onboard JWST \citep{greene2016}. Although we find that the feature amplitude appears to increase with increasing $T_{eq}$, this is partly a consequence of how our model planets' radii scale with $T_{eq}$, which is in accordance with the observed hot Jupiter population \citep{thorngren2016}. In other words, as our $T_{eq}$ = 2100 K model planet is also the largest, with a radius twice that of Jupiter, the spectral features in the planet flux (and thus planet-to-star flux ratio) are also the largest. However, the extent to which the cloud lies above the majority of gas absorption is also important. This suggests that the ideal target for detecting nightside cloud absorption is one with the largest planet-to-star flux ratio and the highest $T_{eq}$ that is also simultaneously cool enough to allow for nightside clouds to persist. 


Our work demonstrates the importance of investigating cloud microphysics across a broad range of planetary parameters and complements similar works with simpler cloud parameterizations but more rigorous 3D atmospheric circulation simulations. Combining these approaches by coupling microphysical models to GCMs and allowing for the advection of cloud particles will be needed to better understand the formation, transport, distribution, and impact of clouds on hot Jupiters. Although several such works already exist \citep[e.g.][]{lines2018neph}, they are computationally expensive and restricted to specific planets, motivating innovation in computational efficiency. In addition, we have only considered variations in $T_{eq}$, but changes in gravity and metallicity beyond 10 m s$^{-2}$ and solar, respectively, could also be important, particularly for the atmospheric thermal structure and mixing strength, both of which strongly impact the cloud distribution. Next generation models that consider a fuller suite of atmospheric processes and a more complete parameter space will be essential as we enter the age of exoplanet atmospheric surveys.

\acknowledgments

We thank V. Parmentier, X. Zhang, M. S. Marley, and J. J. Fortney for enlightening discussions. We thank H. Zhang and W. Z. Gao for their loving support during the writing of this paper. P. Gao acknowledges support from the 51 Pegasi b Fellowship funded by the Heising-Simons Foundation, and NASA through the NASA Hubble Fellowship grant HST-HF2-51456.001-A awarded by the Space Telescope Science Institute, which is operated by the Association of Universities for Research in Astronomy, Inc., for NASA, under contract NAS5-26555. D. Powell acknowledges support from the Ford Foundation Dissertation Year Fellowship Program.



\vspace{5mm}


\bibliography{references}
\bibliographystyle{aasjournal}

\end{document}